\documentclass[12pt, preprint]{aastex}
\doublespace
\usepackage{emulateapj5}
\usepackage{graphicx}
\def\be{\begin{equation}}
\def\ee{\end{equation}}
\def\ba{\begin{eqnarray}}
\def\ea{\end{eqnarray}}

\newenvironment{inlinefigure}{
\medskip
\def\@captype{figure}
\noindent\begin{minipage}{0.999\linewidth}\begin{center}}
{\end{center}\end{minipage}\medskip}

\begin{document}
\title{Locking of the Rotation of Disk-Accreting Magnetized Stars}

\author{M. Long, M.M. Romanova, R.V.E Lovelace}
\affil{Department of Astronomy,
Cornell University, Ithaca, NY 14853\\
Email: Long@astro.cornell.edu}
\singlespace

\begin{abstract}

\doublespace

We investigate the rotational equilibrium state of a disk
accreting magnetized stars using axisymmetric magnetohydrodynamic
(MHD) simulations. In this ``locked'' state, the spin-up torque
balances the spin-down torque so that the net average torque on
the star is zero. We investigated two types of initial conditions,
one with a relatively weak stellar magnetic field and a high
coronal density, and the other with a stronger stellar field and a
lower coronal density. We observed that for both initial
conditions the rotation of the star is locked to the rotation of
the disk. In the second case, the radial field lines carry
significant angular momentum out of the star. However, this did
not appreciably change the condition for locking of the rotation
of the star. We find that in the equilibrium state the corotation
radius $r_{co}$ is related to the magnetospheric radius $r_A$ as
$r_{co}/r_A\approx 1.2-1.3$ for case (1) and $r_{co}/r_A\approx
1.4-1.5$ for case (2). We estimated periods of rotation in the
equilibrium state for classical T Tauri stars, dwarf novae and
X-ray millisecond pulsars.

\end{abstract}


\normalsize

\section{Introduction}

Disk accretion to a rotating star with a dipole magnetic field is
important in a number of astrophysical objects, including T Tauri
stars (Camenzind 1990; K\"{o}nigl 1991), X-ray pulsars (Nagase 1989;
Bildsten et al. 1997), and cataclysmic variables (Warner 1995). An
important property of this interaction is the disruption of the disk
at the Alfv\'{e}n radius, $r_A$, and the ``locking" of the star's
angular rotation at an angular velocity, $\Omega_{eq}$, which is
expected to be of the order of the disk rotation rate at the
Alfv\'{e}n radius, $\Omega(r_A)=(GM/r_A^3)^{1/2}$. However, the
exact conditions for locking and for the value of the equilibrium
rotation rate $\Omega_{eq}$ (when the star does not spin up or spin
down) were not been established.

One of the complicated aspects of the disk-magnetosphere interaction
is the process of angular momentum transport between the disk and
the magnetized star. In the first models of the disk-magnetosphere
interaction it was proposed that the magnetic field has a dipole
configuration everywhere and that the net change between spin-up
torque which arises from the magnetic connection of the star to the
disk within the corotation radius $r_{co}$ and the spin-down torque
which arises from the connection beyond the corotation radius
determines the spin evolution of the star (Ghosh, Lamb \& Pethick
1977; Ghosh \& Lamb 1978, 1979 - hereafter GL79). The corotation
radius is the radius where the disk rotates at the same speed as a
star, $r_{co}=(GM/\Omega_*^2)^{1/3}$. It was suggested that for a
particular value of the star's rotation rate, $\Omega_{eq}$, the
positive spin-up torque balances the negative spin-down torque and
the star is in the rotational equilibrium state (Ghosh \& Lamb 1978,
1979; Wang 1995).

Recent studies of the evolution of the magnetic field threading the
disk and the star led to understanding that the field tends to be
inflated and possibly opened due to the difference in the angular
velocities of the foot-points (Lovelace, Romanova \&
Bisnovatyi-Kogan 1995 - hereafter LRBK; Shu et al. 1994; Bardou
1999, Uzdensky, K\"{o}nigl \& Litwin 2002). In this case a star may
lose some angular momentum through the open field lines and the
equilibrium state will be determined by the balance between
processes of spin-up/spin-down associated with the
disk-magnetosphere interaction, and spin-down associated with the
open field lines.

In some models it was suggested
that the angular momentum transport between the star and the disk
may be much less efficient (Agapitou \& Papaloizou 2000) if the
field lines are opened as proposed by LRBK.
Under such conditions
the rotational equilibrium state will be quite different (e.g.,
Matt \& Pudritz 2004, 2005).
The goal of this paper is to derive the
conditions for the rotational equilibrium state using axisymmetric
MHD simulations of the disk-magnetosphere interaction.

The properties of the rotational equilibrium state depend on the
configuration of the magnetic field threading the star and the disk.
Consequently, analysis of this problem requires two or three
dimensional simulations.
Axisymmetric simulations have
shown that the field lines do open (Hayashi et al. 1996; Miller \&
Stone 1997; Hirose et al. 1997; Romanova et al. 1998; Fendt \&
Elstner 1999).
However, longer runs have shown that the innermost
field lines reconnect to form a closed magnetosphere, and some of
them open and close again in a recurrent manner (Goodson \& Winglee
1997; Goodson, Winglee \& B\"{o}hm 1999; Romanova et al. 2002,
hereafter RUKL; Romanova et al. 2004; Kato et al. 2004, Von Rekowski
\& Brandenburg 2004).
Detailed simulations of the slow, viscous disk
accretion to a rotating star with an aligned dipole field (RUKL)
have shown that on long time-scales, the magnetic field lines in the
vicinity of the Alfv\'{e}n radius $r_A$ are closed or only partially
open, and these lines are important for the angular momentum
transport between the star and the disk.
The balance
between the magnetic flux in closed and open field lines is clearly
important for determining the rotational equilibrium state.

In RUKL, a preliminary search for the conditions of torqueless
accretion was performed and the torqueless accretion was shown to
exist.
In this paper, we give a detailed study of the conditions
for torqueless accretion using an improved axisymmetric MHD code
which makes possible longer simulation runs compared to RUKL.
The main question is: What is the angular rotation rate of a star
$\Omega_{eq}$ for the torqueless accretion given the star's mass
$M$, magnetic moment $\mu$, and accretion rate $\dot{M}$.
Equivalently, if we know the Alfv\'{e}n radius $r_A$, then what is
the corotation radius $r_{co}$ in the rotational equilibrium state?
In earlier theoretical models, it was estimated that the critical
fastness parameter $\omega_s=\Omega_*/\Omega_K(r_A)
=(r_A/r_{co})^{3/2}$ of the equilibrium state is in the range of
$0.47-0.95$ (Li \& Wickramasinghe 1997). This corresponds to the
ratio $r_{co}/r_A \sim 1.1-1.7$.

In this paper we determine the value of $r_{co}/r_A$ for torqueless
accretion numerically by performing a large set of numerical
simulations for different values of $\Omega_*$, $\mu$, $\dot{M}$,
and disk/corona parameters.
We also investigated the
matter flow and the angular momentum transport in the equilibrium
state.
In \S 2, the numerical model is described. We
focus on conditions where the torqueless accretion occurs and study
its dependence on different physical parameters for different cases
in \S 3 and 4.
In \S 5, we apply our results to relevant objects,
such as CTTSs, cataclysmic variables,
and X-ray millisecond pulsars.
We discuss results in Section 6.

\section{Theoretical Ideas and Numerical Simulations}

\begin{inlinefigure}
\centering
\includegraphics[width=3.25in]{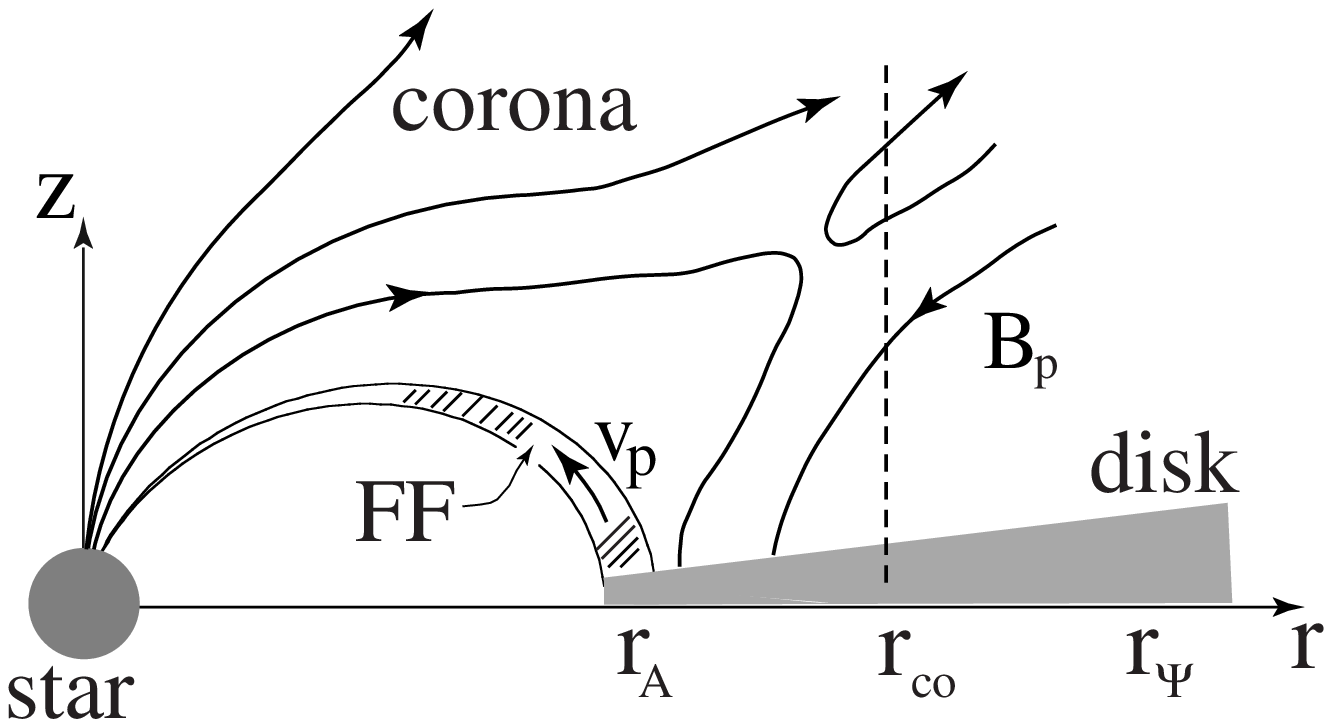}
\figcaption{Sketch of disk accretion to a star with an aligned
dipole magnetic field, where $r_A$ is the Alfv\'{e}n radius,
$r_{co}$ is corotation radius, $r_{\Psi}$ is the deviation radius
where a significant deviation from Keplerian rotation occurs for
$r<r_{\Psi}$ due to the magnetic force on the disk. The dashed line
divides the rapidly rotating region and the slowly rotating region.
FF denotes the funnel flow.}
\end{inlinefigure}

Theoretical aspects of the disk-magnetoshpere interaction were
developed in the 1970s (e.g. Pringle \& Rees, 1972; Ghosh, Lamb \&
Pethick 1977; GL79). As sketched in Figure 1, the initial Keplerian
accretion disk is threaded by the star's magnetic field. The inner
radius of the disk, referred to as the Alfv\'{e}n radius $r_A$,
occurs at the radial distance where the energy-density of
matter in the disk equals to magnetic energy-density, or where the
paramter
$\beta=\rho v^2/(B^2/8\pi) =1$.
(e.g., Lamb, Pethick \& Pines 1973; Davidson \& Ostriker 1973).
We observed from many
simulations that the inner radius of the disk coincides with
this radius. The Alfv\'en radius is usually
derived analytically using a number of approximations.
For a dipole magnetic
field of the star,
$\mu{\equiv}B_*R_*^3=Br^3$,
\begin{equation}
r_A= k_A r_A^{(0)}=k_A{\mu^{4/7}\over ({2GM})^{1/7}\dot{M}^{2/7}}~.
\end{equation}
Here, $G$ is the gravitational constant, $\dot{M}$ is the mass
accretion rate, $M$ is the mass of the star, $r_A^{(0)}$ is the
Alfv\'{e}n radius for the spherical accretion (Elsner \& Lamb 1977).
The coefficient $k_A$ accounts for the fact that the disk accretion
is different from spherical accretion. Using values of the different
parameters obtained from our simulations we obtained $r_A^{(0)}$ and
then compared it with $r_A$. From this we obtained $k_A\approx 0.5$.

In a somewhat different approach, Wang (1995) suggested that the
disk stops or disrupts at the radius $r_0$ where the magnetic and
matter stresses are equal,
$(-r^2B_{\phi}B_{z})_{0}=\dot{M}[\mathrm{d}(r^2\Omega)/\mathrm{d}r]_{0},
$ where the $0-$subscripts indicate evaluation at $r=r_0$,
$B_{\phi}$ and $B_z$ are the components of the magnetic field,
$\Omega$ is the angular velocity of the disk. We took values of
parameters from our simulations and derived the radius the radius
$r_0$. We observed that this radius approximately coincides with
$r_A$.

Whether a star spins up or spins down is due to angular momentum
flow to or out of the star transported by the matter flow and the
magnetic field. Some field lines of the star thread the disk and
transport angular momentum between the disk and the star (e.g.,
GL79). Other field lines are inflated and become open, that is, not
connected to the disk. These open field lines may transport angular
momentum away from the star to the corona without direct interaction
with the disk (Agapitou \& Papaloizou 2000; Matt \& Pudritz 2004).
We observed that both closed and open field lines contribute to the
angular momentum transport to the star.

For the case where the poloidal field lines connect the star and the
disk, the field lines passing through the disk within the corotation
radius $r_{co}$ spin up the star, while the field lines passing
through the disk at large distances spin it down (GL79). The dashed
vertical line in Figure 1 at $r_{co}$ divides the regions which
provide positive and negative torques. There is also a region
$r\lesssim r_{\Psi}$ where the magnetic interaction modifies the
Keplerian rotation of the disk, and where the angular momentum
transport between the star and the disk is important (GL79, RUKL).
If some of the magnetic field lines are open, then angular momentum
may be also transported outward from the disk and from the star
through twisting of open magnetic field lines. Thus part of the
angular momentum flux may be transported by the matter flow, part by
the open field lines, and part by closed field lines connecting the
star and the disk.

\subsection{Main Equations}

The disk-magnetoshpere interaction is considered to be described
by the MHD equations,
\begin{equation}
{\partial \rho \over \partial t}
+\nabla\cdot(\rho\mathbf{v})=0~,
\end{equation}
\begin{equation}
\rho {\mathrm{d}{\bf v} \over \mathrm{d}t} =-\nabla{p}+\rho~{\bf g}
+\frac{1}{4\pi}(\nabla\times{\bf B})\times{\bf B} +{\bf F}^{vis}~,
\end{equation}
\begin{equation}
\frac{\partial\mathbf{B}}{\partial{t}}=
\nabla\times(\mathbf{v}\times\mathbf{B})~,
\end{equation}
\begin{equation}
\frac{\partial(\rho{S})}{\partial{t}}+
\nabla\cdot(\rho\mathbf{v}S)=0~,
\end{equation}
\begin{equation}
\nabla\cdot{\bf B} =0~.
\end{equation}
Here, $S=p/\rho^{\gamma}$ is the entropy per gram, $\mathbf{g}$ is
the gravitational acceleration, ${\bf F}^{vis}$ is the viscous
force. The viscosity model of Shakura and Sunyaev (1973) is used
with viscosity coefficient $\nu=\alpha{c_s^2}/\Omega_K$, where
$c_s=(\gamma p/\rho)^{1/2}$ is the sound speed, and parameter
$\alpha$ is dimensionless with values $\alpha\sim 0.01-0.03$
considered here.

Equations 2-6 were solved with a Godunov-type numerical code
developed by Koldoba, Kuznetzov \& Ustyugova (1992) (see also
Koldoba \& Ustyugova 1994; Ustyugova et al. 1995). We used an
axisymmetric spherical coordinates with a grid $N_r\times
N_\theta=131\times51$. Test simulations with larger or smaller
grids were also performed.

\subsection{Reference Units}

We let $R_0$ denote a reference distance scale, which is equal to
$R_*/0.35$ where $R_*$ is the radius of the star. The subscript
``0" denotes reference values for units. We take the reference
value for the velocity, $v_0=(GM/R_0)^{1/2}$, the angular speed
$\Omega_0=v_0/R_0$, the timescale $P_0=2\pi R_0/v_0$, the magnetic
field $B_0=B_*(R_*/R_0)^3$, the magnetic moment $\mu_0=B_0R_0^3$,
the density $\rho_0=B_0^2/v_0^2$, the pressure $p_0=\rho_0v_0^2$,
the mass accretion rate $\dot{M}_0=\rho_0v_0R_0^2$, and the
angular momentum flux $\dot{L}_0=\rho_0v_0^2R_0^3$. The
dimensionless variables are $\widetilde{R}=R/R_0$,
$\widetilde{v}=v/v_0$, $\widetilde{T}=T/P_0$ etc. In the following
we use these dimensionless units but the tildes are implicit.

\subsection{Boundary and Initial Conditions}

{\it Boundary conditions:} At the inner boundary $R=R_{*}$,
``free" boundary conditions are applied for the density, pressure,
entropy, velocity, and $\phi-$component of the magnetic field,
${\partial\rho}/{\partial R}=0$, ${\partial p}/{\partial R}=0$,
${\partial S}/{\partial R}=0$, ${\partial ({\bf v} - {\bf
\Omega}\times {\bf R})_R}/{\partial R}=0$, ${\partial(R
B_\phi)}/{\partial R}=0$. The poloidal components $B_R$ and
$B_\theta$ are derived from the magnetic flux function
$\Psi(R,\theta)$, where $\Psi$ at the boundary is derived from the
frozen-in condition ${{\partial \Psi}/{\partial t}} + {\bf
v}_p\cdot{\bf \nabla}\Psi = 0$. At the outer boundary, free
boundary conditions are taken for all variables. The outer
boundary was placed far away from the star in order to diminish
the possible influence of this boundary (Ustyugova et al. 1999).

{\it Initial conditions:} The star has a fixed aligned dipole
magnetic field with magnetic moment $\mu$. The initial accretion
disk extends inward to an inner radius $r_d$, and it has a
temperature $T_d$ which is much less than the temperature of the
corona $T_c$. The initial disk and corona are in pressure balance.
The corona above the disk rotates with the angular velocity of the
disk in order to avoid a rapid initial shearing of the magnetic
field. The initial density distributions in the disk and corona
are constrained by the condition that there is a balance of
pressure gradient, gravitational and centrifuge forces. This
balance is necessary in order to have smooth accretion flow which
evolves on the viscous time scale of the disk (see details in
RUKL). To avoid a very strong initial interaction of the disk with
the magnetic field, we take the initial inner radius of the disk
at $r_d=6$ or $r_d=5$, far away from the star.

In order to investigate the rotational equilibrium states, we
performed simulations with two types of initial conditions, type I
and type II. For type I initial conditions we have a magnetic
field which is relatively weak, $\mu=2$, the disk is relatively
thick with fiducial density $\rho_d=1$, and relatively high
coronal density, $\rho_{cor}=0.005$. The initial inner disk radius
is $r_d=6$. The outer radius of simulation domain corresponds to
18$R_0$, that is, about 54$R_*$. Results for this case are
described in \S 3. For type II initial conditions the star's
magnetic field is much stronger, $\mu=10$, the coronal density is
much lower $\rho_{cor}=0.001$, the disk is thinner, and the
initial inner disk radius is $r_d=5$. The outer radius of
simulation domain corresponds to 68$R_0$, that is, about
136$R_*$.These initial conditions, which are more favorable for
inflation and opening of the coronal field lines, are described in
\S 4.


\section{The Equilibrium State of Disk Accretion for
Initial Conditions of Type I}

\subsection{Search for Rotational Equilibrium State}

Whether the star spins up or spins down is determined by the net
flux of angular momentum to the surface of the star, $\dot{L}$.
This flux is composed of two parts, the flux carried by the matter
$\dot{L}_m$, and that carried by the magnetic field, $\dot{L}_f$:
\begin{equation}
\dot{L}=\dot{L}_m+\dot{L}_f~,
\end{equation}
\begin{equation}
\dot{L}_m=-\int \textrm{d}{\bf S} \cdot\rho{r}v_\phi{\bf v}_p~,
\end{equation}
\begin{equation}
\dot{L}_f={1 \over 4\pi}\int \textrm{d}{\bf S}\cdot r B_\phi{\bf
B}_p~,
\end{equation}
where the $p-$subscript denotes the poloidal component, and
$\textrm{d}{\bf S}$ is the outward pointing surface area element of
the star. We performed a set of simulations for different angular
velocities of the star, $\Omega_*$, to find the critical value of
$\Omega_*$, corresponding to the rotational equilibrium state, that
is, the state when $\dot{L}\approx 0$. Other parameters ($\dot{M},
\mu$) were fixed.

We observed that $\dot{L}_f$ is always dominant over $\dot{L}_m$ for
all cases (as in RUKL) and thus we compared $\dot{L}_f$ for these
cases. This was predicted in earlier theoretical research (e.g.,
GL79). We narrowed the set of $\Omega_*$ values so that $\dot{L}_B$
was very small on average (see Figure 2a). Among these we picked the
one for which $\dot{L}_f\approx 0$ that corresponds to the
rotational equilibrium state, $\Omega_*=\Omega_{eq}$. For this case,
the corotation radius $(r_{co})_{eq}\approx 1.7$.

In addition, we calculated matter flux to the star,
\begin{equation}
\dot{M}=-\int \textrm{d}{\bf S} \cdot\rho\mathbf{v}_p.
\end{equation}

Figure 2b shows the mentioned fluxes in the rotational
equilibrium state. Note that this state is typically
reached at $T>50$ because initially the disk is far from the
magnetosphere and it takes time for the disk to move inward.
One can see that the angular momentum flux carried by the field
$\dot{L}_f$ fluctuates around zero.

\begin{inlinefigure}
\centering
\includegraphics[width=3.25in]{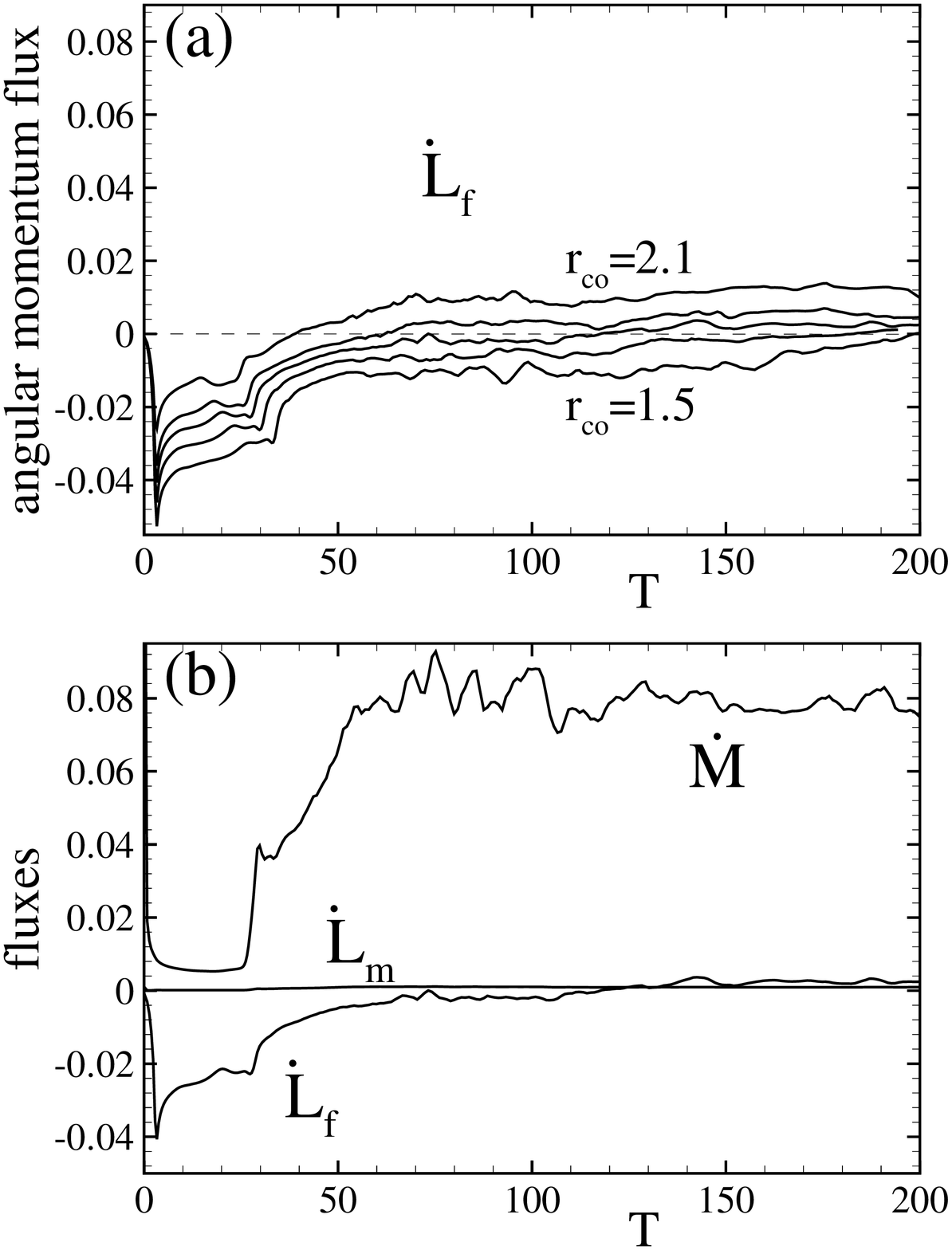}
\figcaption{(a) The evolution of $\dot{L}_f$ for different
corotation radii from $r_{co}=1.5$ (bottom line) to $r_{co}=2.1$
(top line); (b) The evolution of $\dot{M}$ and the angular momentum
fluxes carried by matter $\dot{L}_m$ and the magnetic field
$\dot{L}_f$ in rotational equilibrium state, $r_{co}\approx1.7$.
Time scale is $P_0$ (see \S2.2).}
\end{inlinefigure}

\subsection{Disk-Magnetosphere Interaction in Equilibrium State}

\begin{figure*}[t]
\epsscale{1.8} \plotone{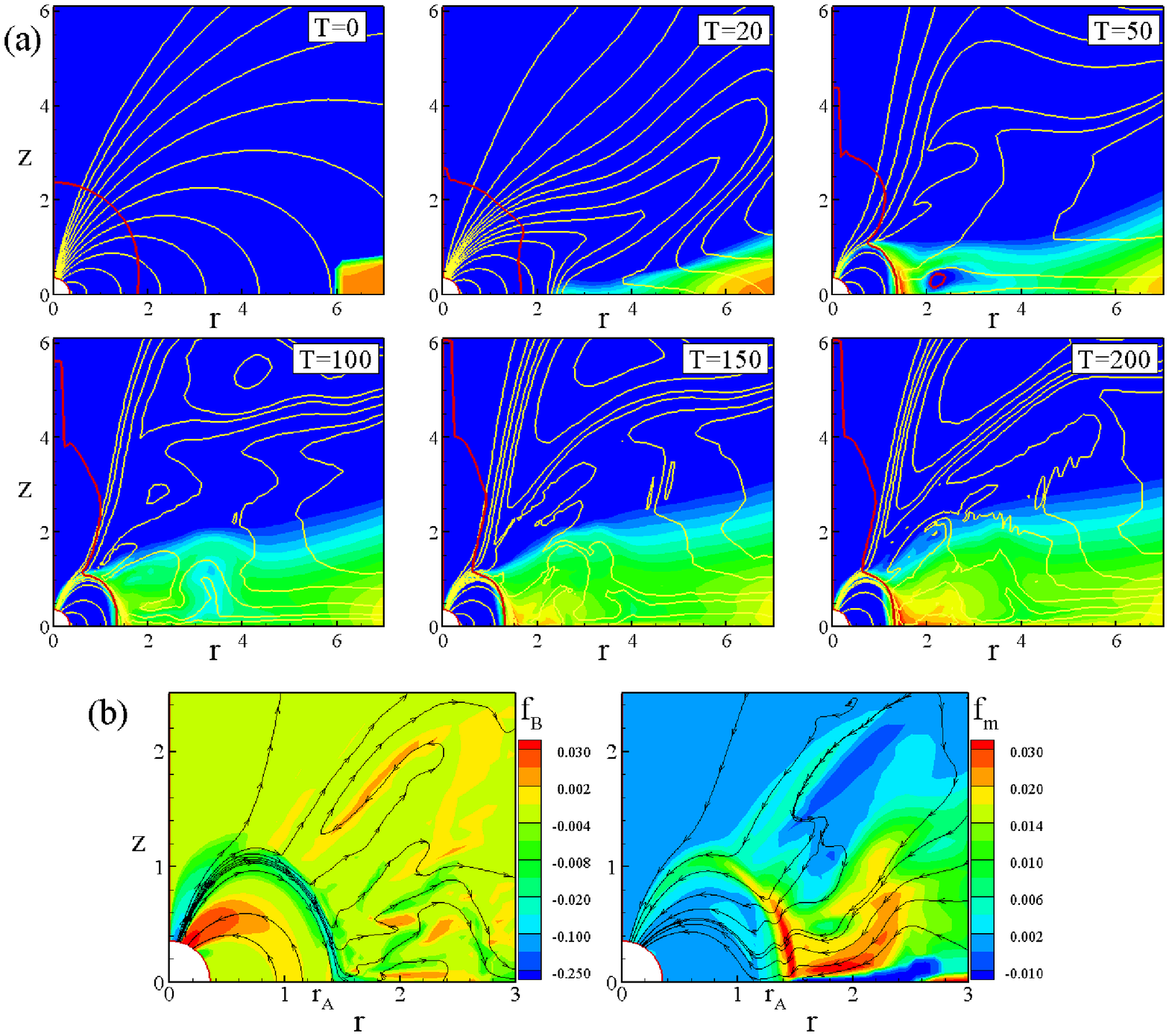} \caption{(a). Evolution of the
density (color background), and the magnetic field (yellow lines)
in rotational equilibrium state for type I initial conditions for
$T=0-200$ rotations. The density changes from $\rho=2$ in the disk
to $\rho=0.005$ in the corona. The bold red line corresponds to
$\beta=1$ (see \S2). (b). Fluxes of angular momentum carried by
magnetic field $\mathbf{f}_B$ (left panel) and by matter
$\mathbf{f}_m$ (right panel) at $T=200$. Color background shows
value of the fluxes, the streamlines with arrows - the direction
of the angular momentum flow.}
\end{figure*}

\begin{figure*}[b]
\epsscale{1.8} \plotone{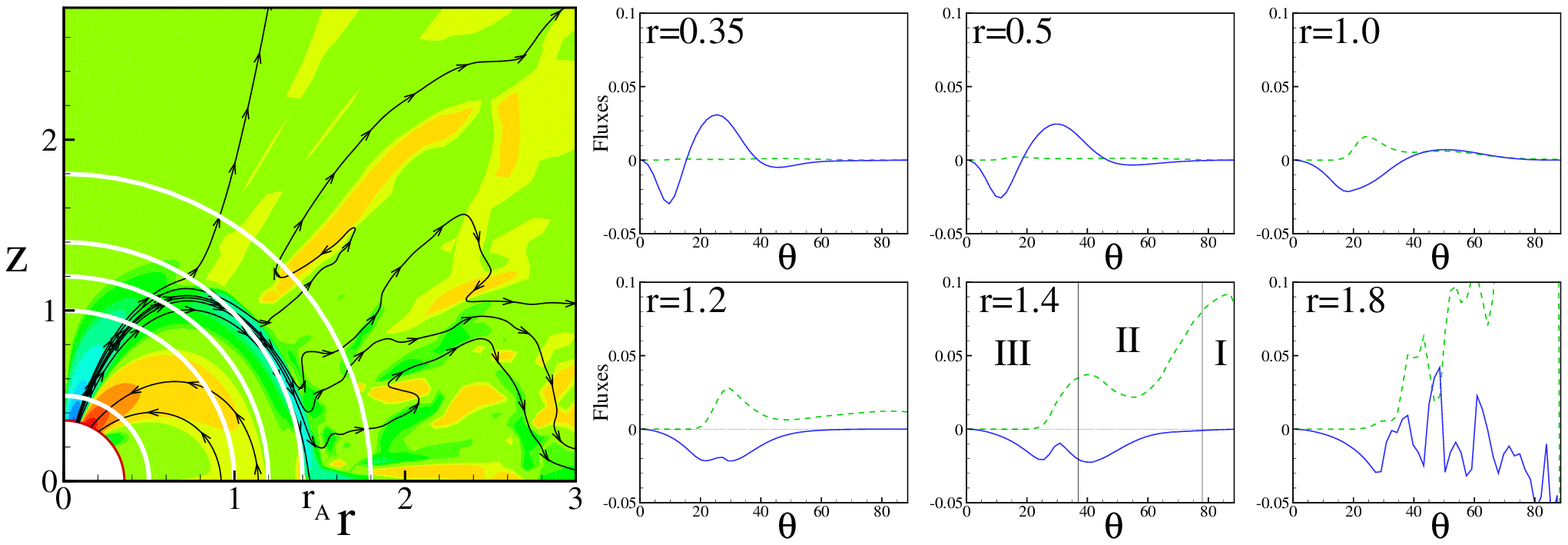} \caption{Angular distribution of
fluxes $F_m(r,\theta)$ and $F_B(r,\theta)$ for type I initial
conditions. The left-hand panel shows radii along which the angular
momentum fluxes are calculated. The right-hand panel shows the
angular momentum fluxes carried by the magnetic field (solid lines)
and the matter (dashed lines) at different radii $r$. The numbers
show the regions where the angular momentum flux carried by the
field is carried mostly by the closed (I), radial (II), or open
(III) field lines.}
\end{figure*}

We now discuss in more detail the rotational equilibrium state where
$r_{co}=1.7$. Figure 3a shows the evolution of matter and the
magnetic flux with time. One can see that initially the disk matter
moves inward, then it stops near magnetosphere and goes to the star
through a funnel flow which is driven up by pressure gradient force
(RUKL;Koldoba et al. 2002). The funnel flow is quasi-stationary
after a time $T\sim 50$. The bold red line divides the regions where
magnetic and matter energy densities dominate, that is, the line
where $\beta=1$ (see \S2). The corresponding Alfv\'{e}n radius is
$r_A\approx 1.2-1.3$. We also estimated the radius $r_0$ derived
from the equality of magnetic and matter stresses. We observed that
this radius is close to $r_A$, $r_0\approx 1.3-1.4$.

Figure 3a shows that the magnetic field for $r>r_A$ is strongly
non-dipolar, and the structure of the field is complicated. The
magnetic field lines initially inflate for $T < 30$; however, later
some of them reconnect forming closed field lines and thus enhancing
the dipole component in the closed magnetosphere. Other field lines,
which are near the axis stay open and represent the lines of
``magnetic tower", which is often observed in different simulations
of the disk-magnetosphere interaction (e.g., Kato et al. 2004;
Romanova et al. 2004). Many field lines are radially stretched by
the accreting matter. These field lines are located in the disk and
above the disk. Most of these field lines are connected to the star.
The field lines above the closed magnetosphere continue to open and
reconnect in quasi-periodic manner. In case of the lower density
corona (see \S 4), inflation is more efficient, and this leads to
larger quasi-periodic oscillations of the magnetic flux and
associated fluxes.

We analyzed the angular momentum transport between the star and the
disk and corona. Figure 3b shows the distribution of the angular
momentum fluxes carried by the field
$\mathbf{f}_B=rB_{\phi}\mathbf{B}_p/4\pi$ (left panel) and carried
by the matter $\mathbf{f}_m=-\rho rv_{\phi}\mathbf{v}_p$ (right
panel). One can see that for $r>r_A$, most of the angular momentum
flux is carried by the matter (see the region with the high positive
angular momentum at the right panel). However, for $r\lesssim r_A$
it is mainly transported by the magnetic field (see left panel of
Figure 3b). The streamlines in the Figure 3b show direction of the
flow of angular momentum. One can see that matter always carries
positive angular momentum towards the star, which tends to spin up
the star. Magnetic field lines carry angular momentum out of the
star through the field lines threading the area of the funnel flow
and corona. The situation with the angular momentum transport seems
to be more complicated compared to one described by GL79.

The angular momentum fluxes change with distance and with angle. We
calculated the angular distribution of fluxes $F_B(r,\theta)=2\pi
r^2\sin\theta\mathbf{f}_B\cdot\hat{r}$ and $F_m(r,\theta)=2\pi
r^2\sin\theta\mathbf{f}_m\cdot\hat{r}$ through the spheres of
different radii $r$. Figure 4 shows that at a large radius, $r=1.8$,
matter carries most of the angular momentum, while the magnetic
field also contributes but with the opposite sign. At smaller radii,
the fluxes become smaller and the largest flux is in the region of
the funnel flow. At the surface of the star, the flux associated
with matter is very small. The flux associated with the field has
two components of the opposite sign which cancel each other
approximately.

The distribution of angular momentum flux is more
complicated compared to that of theoretical models.
However, the equilibrium state does exist and the
ratio $r_{co}/r_A$ is not very much larger than unity. This means
that the rotation of the star is efficiently locked to the rotation
of the inner regions of the disk.

One can see from Figure 3a, that a significant part of the disk is
disturbed by the disk-magnetosphere interaction. Figure 5 shows the
distribution of the angular velocity of the disk in the equatorial
plane. We observed that the angular velocity varies around the
Keplerian value and is usually slightly smaller then Keplerian. The
magnetic field is strongly wound up in the disk so that the
azimuthal field dominates. The inner regions of the disk $r\lesssim
r_{\Psi}\sim 4$ are appreciably influenced by the magnetic field.

\subsection{Dependence on $\mu$ and $\alpha$}

Next, we took a star in the rotational equilibrium state
($r_{co}=1.7$) and varied the magnetic moment of the star $\mu$.
Figure 6a shows that when $\mu$ increases, the rate of change of
angular momentum of the star becomes more negative, that is a star
spins-down. This is connected with the fact that for larger $\mu$,
the magnetosphere is larger and the flux of the positive angular
momentum to the star is smaller than in the equilibrium state. For
example, for $\mu=4$, the inner disk radius is at $r_A\approx 1.6$.
At this inner radius, the disk rotates much slower than the star and
a star strongly spins down. Roughly, the dependence is
$\dot{L}_f\sim-0.02\mu+0.04$. The balance between spin-up and
spin-down torques is similar to that suggest by the GL79 model,
where positive and negative torques are associated with regions
within and beyond the corotation radius. We also observed different
regions contributing positive and negative angular momentum fluxes.
However now the distribution of the external to the Alfv\'{e}n
surface field lines is more complicated.

We calculated the average value of $r_A$ for each $\mu$ and obtained
the dependence $r_A\sim\mu^{\kappa}$, with $\kappa=0.36$. This
coefficient $\kappa$ is somewhat different from that of equation (1)
where $k=4/7 \approx 0.57$. The difference may be connected with the
fact that in the theoretical analysis the magnetic field is assumed
to be a pure dipole field everywhere, whereas the simulations show
that the actual field is different from a dipole for $r\gtrsim r_A$.
In this and all above simulations, the viscosity coefficient was
fixed at the value $\alpha=0.02$.

We also performed simulation runs for different values of the
disk viscosity, $\alpha=0,~ 0.01,~ 0.02,$ and $ 0.03$, with other
parameters fixed.
We observed that for larger $\alpha$ the
accretion rate is larger, and the Alfv\'{e}n radius $r_A$ is
smaller. Thus, the region of positive torque becomes larger than the
region of negative torque, and $\dot{L}_f$ is larger.
In other words, at a higher accretion rate
$\dot{M}$, incoming matter brings more of positive angular momentum,
which is transferred to larger angular momentum flux carried by the
field $\dot{L}_f$. The angular momentum flux increases with $\alpha$
(see Figure 6b) as about $\dot{L}_f\sim\alpha^{1.9}$ for $\alpha\geq
0.01$.

\begin{inlinefigure}
\centering
\includegraphics[width=3.25in]{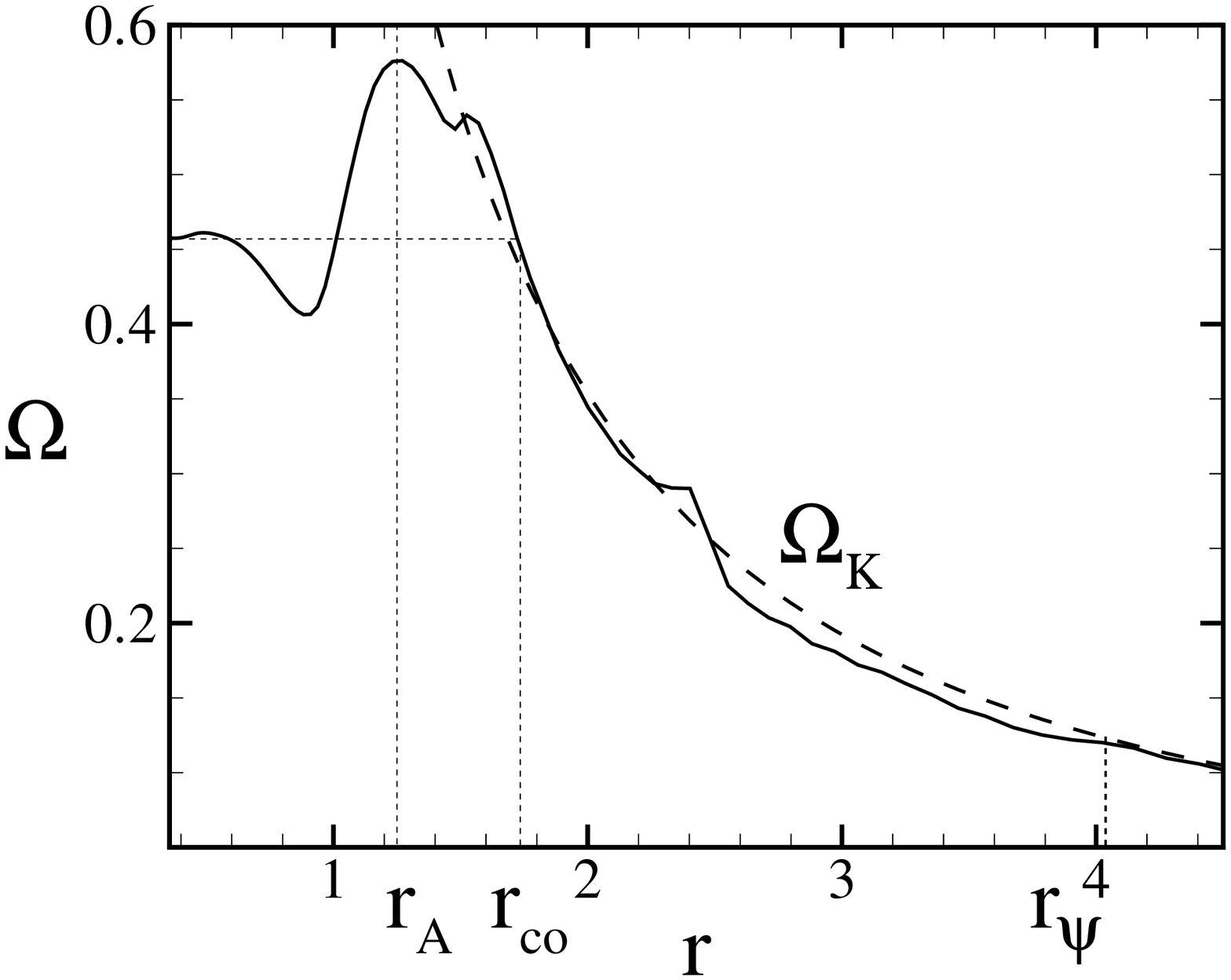}
\figcaption{Radial distribution of angular velocity of the disk
along in the equatorial plane at $T=180$. The dash line shows
Keplerian angular velocity.}
\end{inlinefigure}

\begin{inlinefigure}
\centering
\includegraphics[width=3.25in]{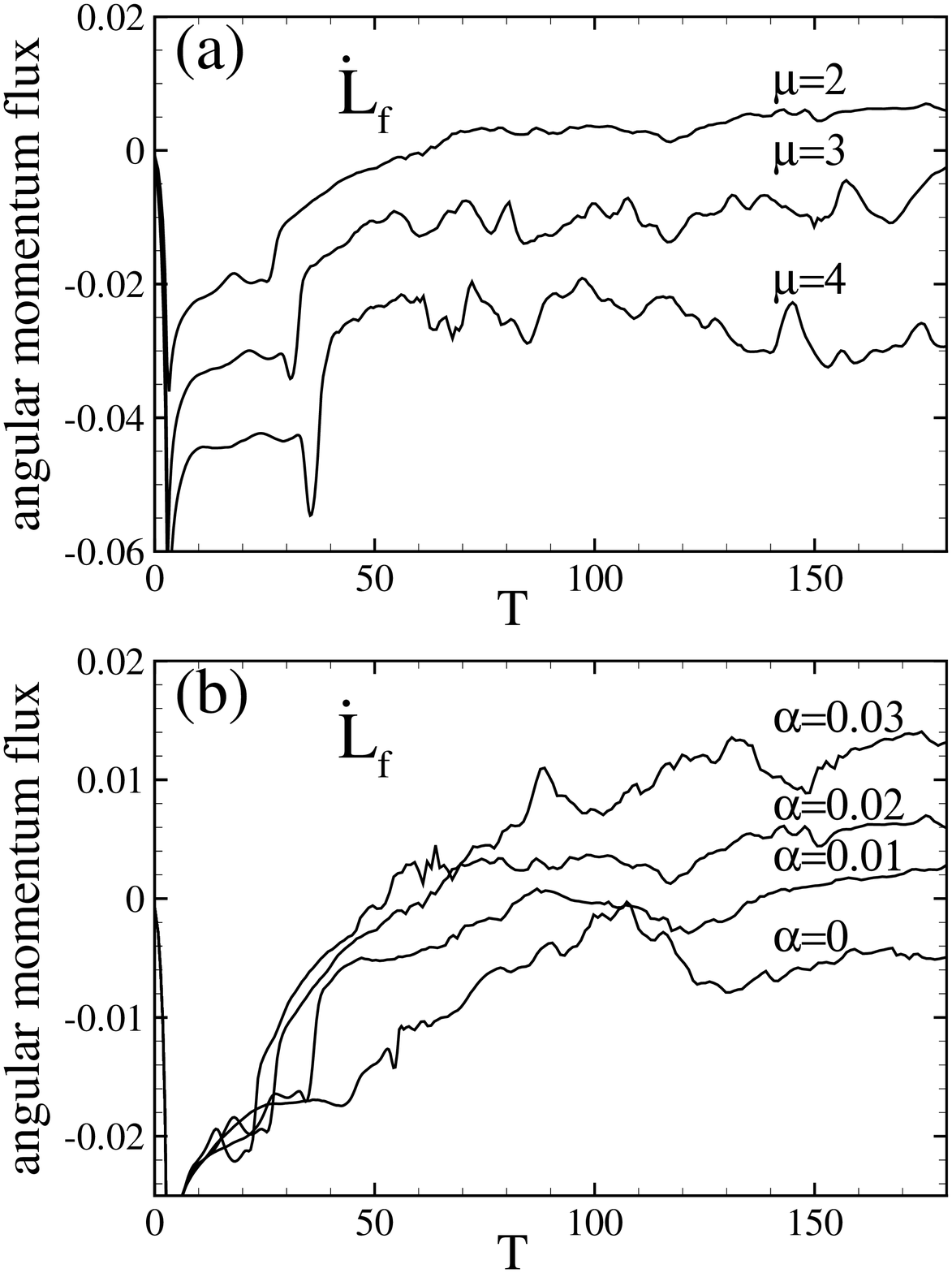}
\figcaption{(a) The evolution of the angular momentum flux to the
star carried by the magnetic field $\dot{L}_f$ for different
magnetic moments of the star $\mu$ with other parameters fixed. (b)
The evolution of the angular momentum flux carried by magnetic field
$\dot{L}_f$ for different values of of the disk viscosity $\alpha$
with other parameters fixed.}
\end{inlinefigure}

\section{Equilibrium State of Disk Accretion for Initial Conditions
of Type II}

\begin{inlinefigure}
\centering
\includegraphics[width=3.25in]{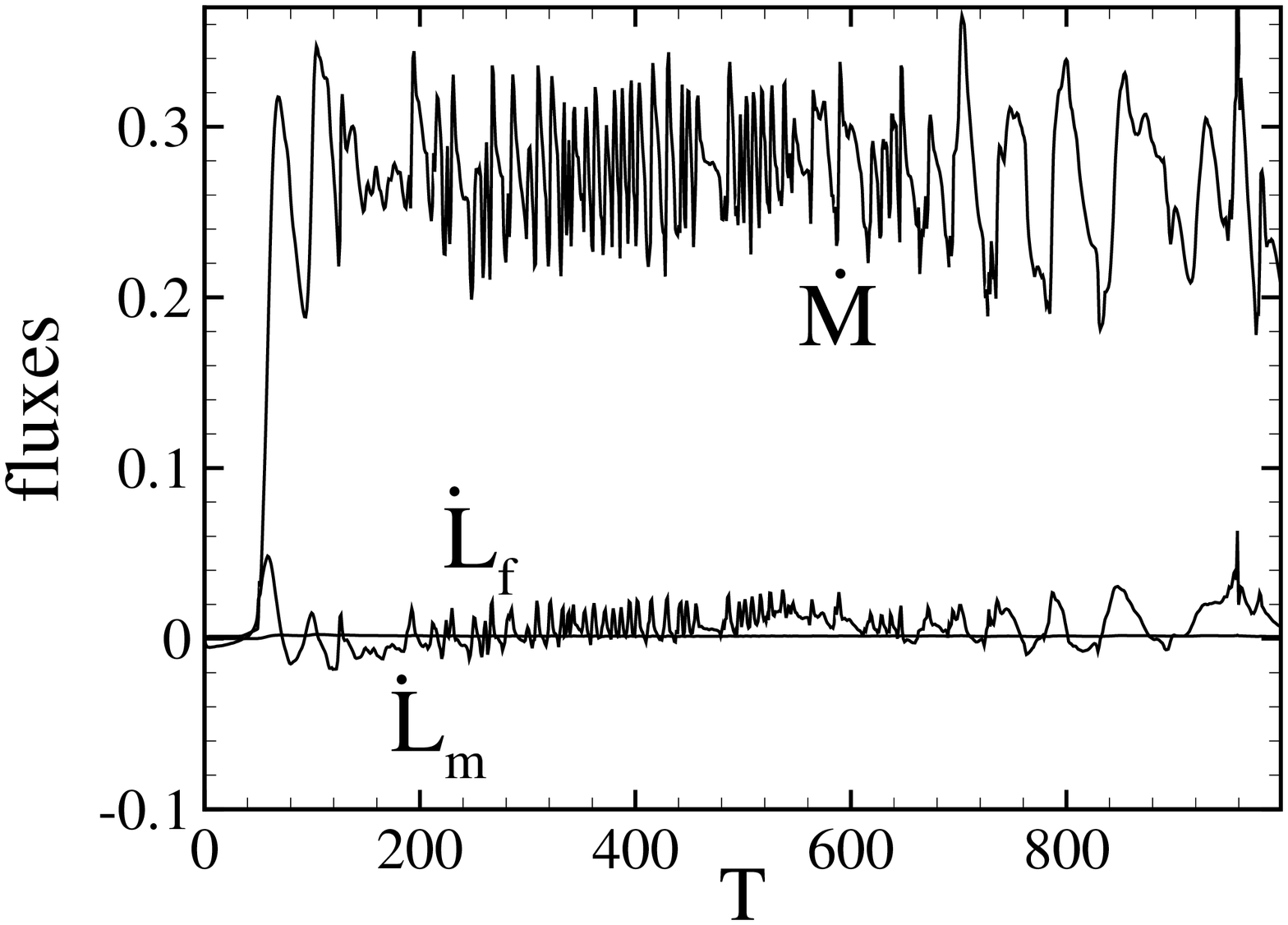}
\figcaption{The evolution of the matter flux $\dot{M}$ and the
angular momentum fluxes carried by matter $\dot{L}_m$ and magnetic
field $\dot{L}_f$ in the rotational equilibrium state for the type
II initial conditions.}
\end{inlinefigure}

\begin{figure*}[t]
\epsscale{1.8} \plotone{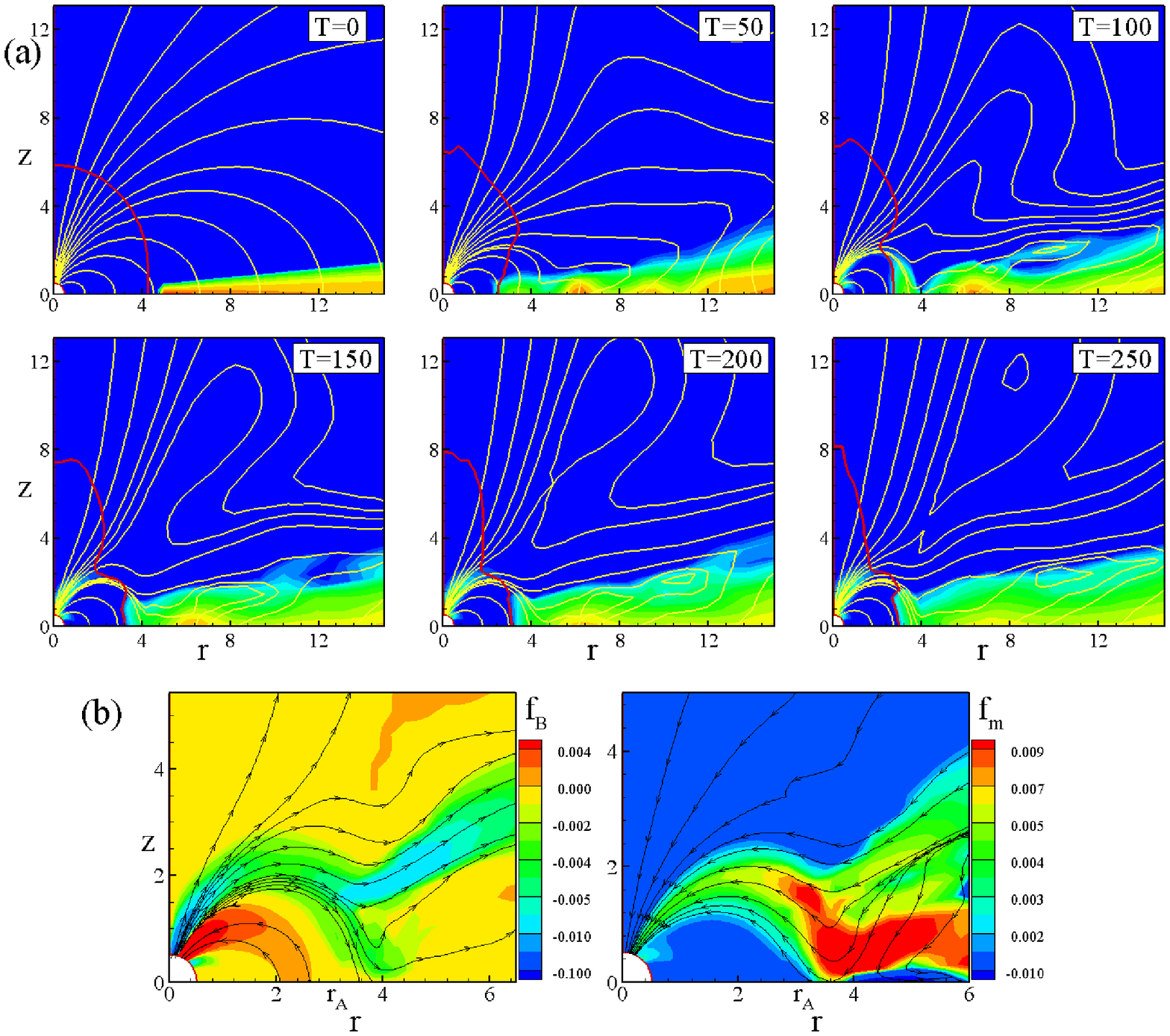} \caption{(a). Evolution of the
density (color background), and the magnetic field lines (yellow
lines) in rotational equilibrium state of type II initial
conditions for $T=0-250$ rotations. The density changes from
$\rho=2$ in the disk to $\rho=0.001$ in the corona. The bold red
line corresponds to $\beta=1$ (see \S2). (b). Fluxes of angular
momentum carried by magnetic field $\mathbf{f}_B$ (left panel) and
by matter $\mathbf{f}_m$ (right panel) at $T=250$. Color
background shows value of the fluxes, the streamlines with arrows
- the direction of the angular momentum flow.}
\end{figure*}

\begin{figure*}[b]
\epsscale{1.8} \plotone{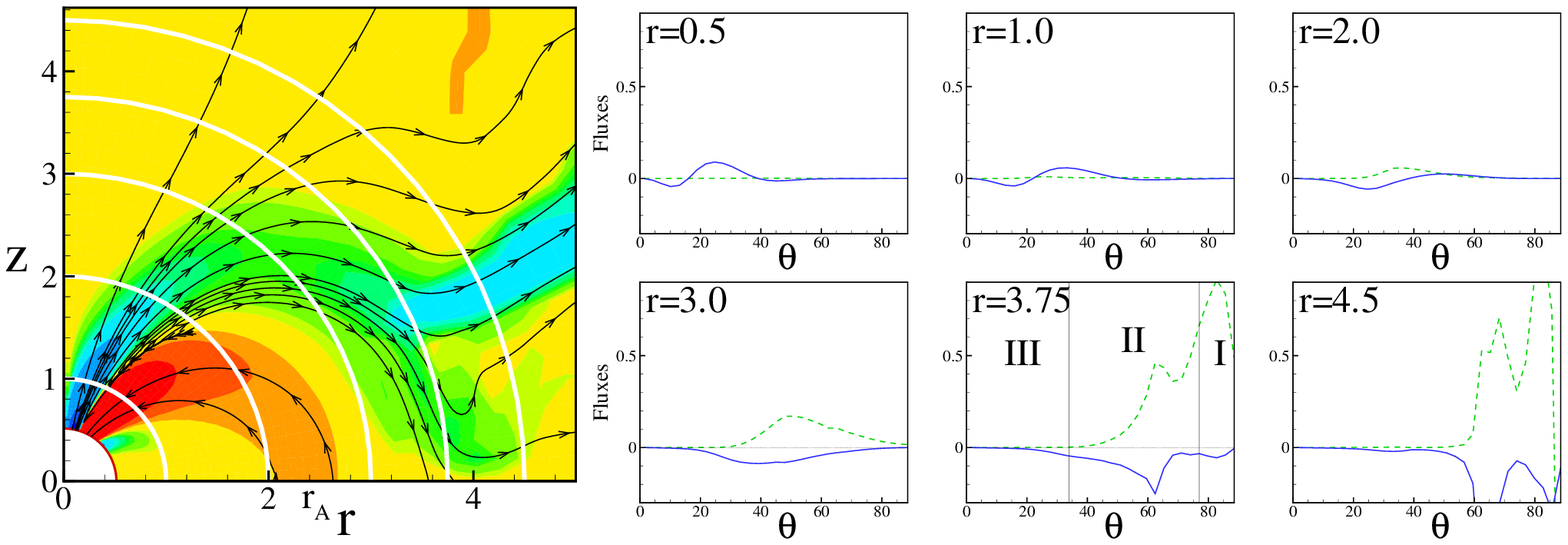} \caption{Angular distribution of
fluxes $F_m(r,\theta)$ and $F_B(r,\theta)$ for initial conditions of
type II. The left-hand panel shows radii along which the angular
momenta are calculated. The right-hand panel shows angular momentum
fluxes carried by magnetic field (solid lines) and matter (dashed
lines. The numbers show the regions where the angular momentum flux
carried by the field is carried mostly by the closed (I), radial
(II), or open (III) field lines.}
\end{figure*}

Here we consider another case corresponding to quite different
initial conditions referred to as type II. This case has a stronger
magnetic field, $\mu=10$, and a much lower coronal density,
$\rho_{cor}=0.001$.
The density in the corona influences the
evolution of the magnetic field in the corona, namely, the inflation
and opening of the magnetic field lines is favored.
As discussed earlier, the angular momentum transport between the
disk and star occurs in part through the magnetic field.
In addition, we took a different
structure of the disk, which is thinner and located initially at
$r_d=5$, which is closer to the star for such a strong field. We
were able to model the stronger magnetic field because we
increased the radius of the star (the inner boundary) to $R_*=0.5$
(versus $R_*=0.35$ in other simulation runs).
In addition, we were able to use a grid $N_R \times
N_\theta = 91\times31$ which speeded up the simulations
and allowed longer runs,
up to $1000~P_0$.

For this case, we also performed a set of simulations for different
corotation radii and found the state which gives torqueless
accretion, $\dot L\approx 0$.
We found that this state corresponds to a corotation radius
$r_{co}\approx 4.6$.
Figure 7 shows the evolution of the fluxes in this case.
The angular momentum flux associated with the
matter is very small as in our main case.
The flux due to the
magnetic field, $\dot{L}_f$, varies
more strongly than in the main
case.
This is because the coronal density
is lower so that variations of the magnetic
field in the corona are larger.
In addition, the accretion rate
$\dot M$ is several times larger
than that for our main case, because the disk is different compared
to the main case.

We analyzed the rotational equilibrium state in detail. Figure 8a
shows the evolution of the disk and magnetic field in the inner part
of the simulation region. We observed that the accretion disk
stopped at much larger distances, $r_A\approx 3.1$, compared to our
main case ($r_A\sim 1.2-1.3$) because of much stronger magnetic
field. This radius again coincides with the $\beta=1$ line (see the
bold red line on figure 8a). We also estimated the radius $r_0$
derived from comparison of magnetic and matter torques. Again, we
obtained similar but a somewhat larger radius, $r_0\approx 3.2$. We
observed that in type II case, the ratio $r_{co}/r_A\sim1.4-1.5$.
This number is slightly larger than that in the main case
($r_{co}/r_A \sim 1.3-1.4$) which may be connected with contribution
of the inflated field lines to the negative torque.

Figure 8b shows the angular momentum fluxes carried by the field
$F_B(r,\theta)$ and by the matter $F_m(r,\theta)$ . One can see that
the disk matter brings positive spin-up torque (right panel), while
magnetic field carries both negative and positive torques (left
panel). In this case a significant part of the spin-down flux is
carried by the inflated and radially stretched field lines. This is
different from the type I initial conditions case.

Figure 9 shows angular distribution of fluxes $F_B(r,\theta)$ and
$F_m(r,\theta)$ along the spheres of different radii $r$ from
$r=0.5$ to $r=4.5$. One can see, that at large distances, $r=4.5$,
there is a large positive angular momentum flux carried by the
matter of the disk, and there is a large negative angular momentum
flux carried by the radially stretched field lines located above the
disk and in the disk. Field lines above the disk are inflated field
lines which connect to the disk at large radii $r>20$. They carry
both, positive matter flux, and negative spin-down flux. In this
region the density is smaller than in the disk, however, the field
lines are strongly wound and azimuthal velocity is large, this is
why both fluxes associated with matter and the field are large in
this region. The plot at $r=3.75$ shows relative input from
different sets of field lines. It shows that in the region of the
closed field lines (region I) matter carries positive torque while
field lines carry only small negative torque. Most of both torque is
carried by the radially stretched field lines (region II), while the
role of vertically inflated field lines of ``magnetic tower" (region
III) is very small. At even smaller radii, $r=3,2,1$ and $0.5$, both
torques become smaller, like in type I case.

\section{Equilibrium State in Applications to Different Stars}

The rotational equilibrium state is expected to be the most probable
state for different disk accreting magnetized stars. These include
Classical T Tauri Stars, cataclysimic variables, and X-ray pulsars.
For example, the slowly rotating CTTSs have spin rates as low as
$\sim10\%$ of breakup speed (Bouvier et al. 1993). The fact that
CTTSs rotate slowly strongly suggests that they are in rotational
equilibrium state.

Below we estimate periods of rotation for different weakly
magnetized stars. Our simulations were performed for cases
$r_A/R_*=4-6$ and can be applied to systems with different scales in
which this ratio is satisfied. Observations show that in CTTS this
ratio is in the range $r_A/R_*=4-8$ (Bouvier et. al., 1993, Edwards
et al., 1993),  or, $r_A/R_*=3-10$ (Kenyon, Yi \& Hartmann, 1996).
Similar or smaller magnetospheres are expected in accreting
millisecond pulsars (van der Klis 2000) and dwarf novae cataclysmic
variables (Warner 2004).

For our example we take $r_A/R_*=4$, take obtained from simulation
ratio $r_{co}=1.4 r_A$ and derive the angular velocity for
rotational equilibrium, $\Omega_{eq}$:
\begin{equation}
\frac{\Omega_{eq}}{\Omega_{*K}}=
\left(\frac{R_*}{r_{co}}\right)^{3/2}\approx0.09~,
\end{equation}
which is about $\sim10\%$ of breakup speed of the star. Here,
$\Omega_{*K}$ is the Keplerian angular speed at $R_*$. The value
of $\Omega_{eq}/\Omega_{*K}$ we find is close to the one observed
in CTTSs.

\subsection{Classical T Tauri Stars (CTTSs)}

The angular velocity of the star can be written as
$\Omega_*=({GM}/{r_{co}^3})^{1/2}$.
Combining this with the rotational
equilibrium condition $r_{co}/r_A\sim1.3-1.5$ (an average 1.4 was
taken here) and using Eqn. 1 for the Alfv\'{e}n radius, we can
get the critical value of angular velocity $\Omega_{eq}$ or
rotation period $P_{eq}=2\pi/\Omega_{eq}$.
For the CTTS we obtain the equilibrium rotation period,
\begin{eqnarray}
P_{eq} \approx 4.6~{\rm days}
\Big(\frac{0.8M_\odot}{M}\Big)^{5/7}
\Big(\frac{10^{-7}M_\odot/\mathrm{yr}}{\dot{M}}\Big)^{3/7}
\times
\nonumber \\
\Big(\frac{B}{10^3\mathrm{G}}\Big)^{6/7}
\Big(\frac{R_*}{2R_{\odot}}\Big)^{18/7},
\end{eqnarray}
where $B$ is the surface magnetic field of the star. Our result is
in the range of an observational bimodal distribution of rotation at
period with peaks near 3 and 8 days (Attridge \& Herbst 1992).
Edwards et al. (1993) proposed that the distribution near 8 days may
be caused by the disk locking. We should note that both peaks may be
connected with the disk locking, but at different parameters, say,
$\dot{M}$ and $B$.

\subsection{Cataclysmic Variables}

We next consider dwarf novae which belong to a subtype of the
cataclysmic variables where the magnetic field is expected to be
small but still possibly dynamically important for the accretion.
The accretion disks are stopped at small radii by the
white dwarf magnetosphere. The accreting material then leaves the
disk and follows the magnetic field lines down to the star's
surface in the vicinity of the magnetic poles.
Taking typical values for the white dwarf and accretion disk, we obtain the
period of these stars in rotational equilibrium state,
\begin{eqnarray}
P_{eq}\approx57~\mathrm{s}\Big(\frac{M_\odot}{M}\Big)^{5/7}
\Big(\frac{10^{-8}M_\odot/\mathrm{yr}}{\dot{M}}\Big)^{3/7} \times
\nonumber \\
\Big(\frac{B}{10^5\mathrm{G}}\Big)^{6/7}
\Big(\frac{R}{7.0\times10^{8}\mathrm{cm}}\Big)^{18/7}~.
\end{eqnarray}
The observed periods of Dwarf Novae Oscillations (DNOs)
are in the range of $7-70$ s (Warner, 2004).

\subsection{Millisecond Pulsars}

Our simulations are also applicable to millisecond X-ray pulsars,
which are the accreting low-magnetic-field neutron stars (van der
Klis, 2000).
The period of such a star in the
rotational equilibrium is
\begin{eqnarray}
P_{eq}\approx1.8~\mathrm{ms}\Big(\frac{1.4M_\odot}{M}\Big)^{5/7}
\Big(\frac{10^{-9}M_\odot/\mathrm{yr}}{\dot{M}}\Big)^{3/7} \times
\nonumber \\
\Big(\frac{B}{10^8\mathrm{G}}\Big)^{6/7}
\Big(\frac{R}{10^{6}\mathrm{cm}}\Big)^{18/7}.
\end{eqnarray}
For all accretion-powered millisecond X-ray pulsars we have known,
the spin frequencies range from 185Hz to 435Hz (Wijnands, R.
2004), or, the spin rate is in range of 2.3-5.2 ms. These values
may be obtained from Eqn.14 at smallere $\dot{M}$ or larger $B$.

\section{Discussion}

We investigated the conditions for the rotational equilibrium
or``torqueless" accretion state using axisymmetric MHD simulations.
In such a state the total angular momentum flux to the star is
approximately zero. We found the equilibrium states by gradually
changing the angular velocity of the star with the other parameters
fixed. We considered two main cases: One with relatively low
magnetic field and dense corona and second with much stronger field
and lower density corona. We observed that in both cases the
rotation of the star is approximately locked to the rotation of the
inner radius of the disk such that a star rotates somewhat slower
than the inner radius of the disk. In the first case the ratio
between corotation radius of the star and the Alfv\'{e}n radius
$r_A$ (where the disk is disrupted) is $r_{co}/r_A\sim 1.3-1.4$. In
the second case, where stronger inflation of the magnetic field was
observed, this ratio is only slightly larger, $r_{co}/r_A\sim
1.4-1.5$. We observed that in the first case, the angular momentum
transport is associated with the closed field lines at the inner
radii of the disk. Open field lines spin down the star, but the role
of this spin-down is small. In the second case outflow of angular
momentum along the inflated field lines is more significant than in
the first case, however this did not change the result, which
probably means that the angular momentum transport associated with
the inner regions of the disk and the region of the funnel flow is
more significant. Thus, in both cases the magnetic interaction
effectively locks the rotation rate of the star to a value which
depends mainly on the mass accretion rate and the star's magnetic
moment. We should note that most of coronal region is still matter
dominated. This is connected with the fact that the magnetic
energy-density of the dipole decreases with distance as $\sim
R^{-6}$ so that it is difficult to set up a magnetically dominated
corona. Goodson et al. (1997, 1999) were able to model such a corona
by arranging fast fall-off of coronal density with the distance.
However, their initial conditions are sufficiently far from
equilibrium that the torqueless accretion was not established.
Future simulations with even lower coronal densities will help to
understand whether a star is always locked to the rotation rate of
the inner radius of the disk. Our simulations for two very different
initial conditions have shown very similar results for the
rotational equilibrium which may be a sign that the disk-locking may
be a similar for all slowly rotating stars.

We applied our simulation results to Classical T Tauri stars, where
disk locking may explain their slow rotation.
Also, we estimated the probable periods of rotation of other
accretion powered systems, such as dwarf novae and X-ray millisecond
pulsars.

\acknowledgments The authors thank Drs. Koldoba and Ustyugova for
developing of codes and helpful discussions. This work was supported
in part by NASA grants NAG 5-13220, NAG 5-13060 and by NSF grant
AST-0307817.

\end{document}